\documentclass[aps,prl,nofootinbib,amsmath,twocolumn,preprintnumbers,superscriptaddress]{revtex4-1}

\usepackage{amssymb,esvect,amsmath,graphicx,latexsym,amsthm,slashed,eso-pic,hyperref}
\usepackage[hang,flushmargin]{footmisc}
\usepackage{adjustbox}
\usepackage{placeins}
\usepackage{bbold}
\usepackage{multirow}
\usepackage{slashed}
\usepackage{cancel}

\usepackage[normalem]{ulem}

\DeclareMathOperator{\cm}{cm}

\DeclareMathOperator{\nm}{nm}

\DeclareMathOperator{\mum}{\mu \mathrm{m}}

\DeclareMathOperator{\esi}{\epsilon_{\mathrm{Si}}}
\DeclareMathOperator{\esio}{\epsilon_{\mathrm{SiO_2}}}
\DeclareMathOperator{\Esio}{\mathit{E}_{\textrm{SiO2}}}

\begin{document}

\title{Dual-sided Charge-Coupled Devices}

\author{Javier Tiffenberg}
\thanks{javiert@fnal.gov}
\affiliation{Fermi National Accelerator Laboratory, PO Box 500, Batavia IL, 60510, USA}

\author{Daniel Ega\~na-Ugrinovic}
 \thanks{\textit{Corresponding author:}~danielegana@gmail.com}
%\email{danielegana@gmail.com}  
\affiliation{Perimeter Institute for Theoretical Physics, Waterloo, ON N2L 2Y5}

\author{Miguel Sofo Haro}
\affiliation{
Fermi National Accelerator Laboratory, PO Box 500, Batavia IL, 60510, USA}
\affiliation{Centro At\'omico Bariloche, CNEA/CONICET/IB, Bariloche, Argentina}

\author{Peizhi Du}
\affiliation{New High Energy Theory Center, Rutgers University, Piscataway, NJ 08854, USA}

\author{Rouven Essig}
\affiliation{C.N. Yang Institute for Theoretical Physics, Stony Brook University, Stony Brook, NY, 11794, USA}

\author{Guillermo Fernandez-Moroni}
\affiliation{Fermi National Accelerator Laboratory, PO Box 500, Batavia IL, 60510, USA}

\author{Sho Uemura}
\affiliation{Fermi National Accelerator Laboratory, PO Box 500, Batavia IL, 60510, USA}

\preprint{YITP-SB-2023-13, FERMILAB-PUB-23-376-PPD}
 \begin{abstract} 
Existing Charge-Coupled Devices (CCDs) operate by detecting either the electrons or holes created in an ionization event. We propose a new type of imager, the Dual-Sided CCD, which collects and measures both charge carriers on opposite sides of the device via a novel dual-buried channel architecture.
We show that this dual detection strategy provides exceptional  dark-count rejection and enhanced timing capabilities. 
These advancements have wide-ranging implications for dark-matter searches, near-IR/optical spectroscopy, and time-domain X-ray astrophysics. 
\end{abstract}
 
\maketitle

\newpage 

\section{Introduction}
Charge-Coupled Devices (CCDs) are broadly used as scientific instruments, offering unsurpassed imaging quality~\cite{6768140,amelio1970experimental,damerell1981charge,janesick1987scientific}. Existing CCDs collect and read only either the electrons or holes created in a bulk ionization event~\cite{janesick1987scientific}, while the charge carriers with opposite polarities are discarded. In this \textit{letter}, we propose a novel device  that enables measuring both charge carriers, the ``Dual-Sided CCD'' (DCCD), and we show that this detector has two advantages over standard single-polarity CCDs: improved rejection of dark counts (DCs) from various sources, including surface and spurious events, and enhanced timing capabilities. These advancements are critical for several scientific applications, including  searches for dark matter
~\cite{Tiffenberg:2017aac,Crisler:2018gci,SENSEI:2019ibb,SENSEI:2020dpa,SENSEI:2021hcn,DAMIC:2016lrs,Castello-Mor:2020jhd,DAMIC:2020cut,DAMIC-M:2023gxo,Du:2020ldo,Oscura:2022vmi,Oscura:2023qik,Oscura:2023qch,Du:2022dxf}, neutrinos~\cite{CONNIE:2019swq,Fernandez-Moroni:2020yyl,Fernandez-Moroni:2021nap}, spectroscopy~\cite{Drlica-Wagner:2020wck,DESI:2022lza}, and time-domain astrophysics~\cite{arnaud2011handbook,uttley2014x,Ingram:2019mna,CACKETT2021102557,Mushotzky:2019lpm,gaskin2019lynx,Feigelson2022,doi:10.1146/annurev-astro-052920-112338}. 

\begin{figure}[h]
\centering
\includegraphics[width=0.95\columnwidth]
{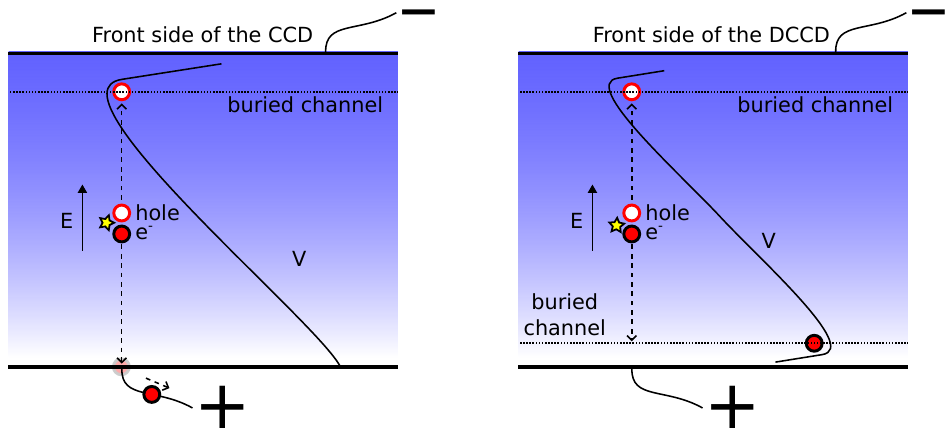}
\caption{Cross-section of a standard p-channel CCD collecting holes in its buried channel at the front of the sensor (left panel) and of a DCCD that collects holes at the front channel and electrons at the back channel (right panel).
}
\label{fig:burried_channel}
\end{figure}

In standard CCDs an event  is detected by establishing an electrostatic potential that drifts the charge towards the ``buried channel'' \cite{boyle1974buried}, a potential extremum located close to one of the detector surfaces  where the charge is stored away from gate insulating layers to avoid traps. The channel can be p-type, in which case it collects holes ($h$), or n-type, for electron ($e$) collection \cite{janesick1987scientific}. To collect \textit{both} positive and negative charge carriers, a DCCD is equipped with  channels of opposite polarities on the detector's front and backsides, as illustrated in Fig.~\ref{fig:burried_channel}.

The advantages of such a design can be now made apparent. As shown in Fig.~\ref{fig:burried_channel}, bulk signal events populate both the front and back channels with opposite-polarity charges. Various DC sources, on the other hand, lead predominantly to \textit{single-channel} events. 
Surface DCs, which are a leading background in state-of-the-art devices 
\cite{Du:2020ldo,Du:2023soy,SENSEI:2019ibb,SENSEI:2020dpa,janesick1987scientific,SENSEI:2021hcn}, occur thermally or from charge leakage in thin regions in-between the buried channels and the surface detector gates where the traps from insulating layers are located~\cite{janesick1987scientific,hynecek1979virtual,saks1980technique,ranuarez2006review,PhysRev.87.835,janesick1987scientific,ranuarez2006review,lenzlinger1969fowler,weinberg1982tunneling,maserjian1974tunneling,srivastava1985electrical}. Given the potential profile in these  regions, such DCs only populate either the front or backside channels with a hole or electron, respectively,  as any charge with the opposite polarity is drifted away towards the gates. Spurious or clock-induced charge generated during charge transfer, a dominant source of DCs in several devices~\cite{janesick1987scientific,SENSEI:2020dpa,jerram2001llccd,daigle2010darkest,daigle2012characterization,bush2021measurement}, and all events that occur in the serial registers, transfer gates, or amplifiers, also lead to single-channel backgrounds. By discriminating such single-channel backgrounds from the dual-channel signals of bulk events, DCCDs can provide strong DC rejection. 

Improved timing is also enabled by the readout of both charges. Standard CCDs have a time resolution that is typically limited by the time it takes to read out all the pixels of the array. In a DCCD this resolution can be significantly improved to be the time needed to read out a \textit{single serial register} (a ``row'' of pixels). In megapixel devices this represents a three-order of magnitude improvement, and leads to time resolutions that can be as low as $\sim$10\,$\mu$s in EMCCDs~\cite{jerram2001llccd} and $\sim$1~s in sub-electron-noise Skipper-CCDs~\cite{SENSEI:2020dpa}.
Enhanced timing is achieved by reading out the device continuously and looking for correlations between the charges collected on the front and back channels. While these correlations can be obscured by event pileup, we show with simulations that this technique can be used in realistic applications.

We organize this letter as follows. We first discuss the basic design of a DCCD, device inversion, and present simulations demonstrating improved DC rejection and timing. We then discuss an example application of the DCCD as a dark matter detector. We conclude by commenting on future  directions.  Natural units $\hbar=c=1$ are used throughout.

\section{Detector architecture}

\subsection{Dual Buried Channel}
We design a DCCD starting from a standard three-phase~\cite{sequin1974charge}, buried channel~\cite{boyle1974buried}, high-resistivity~\cite{holland1989fabrication,holland1996200,holland2003fully} p-channel~\cite{spratt1997effects,hopkinson1999proton,bebek2002proton} Si CCD.\footnote{A  design starting from an n-channel CCD can be obtained by inverting p and n-type materials.} In a p-channel CCD, a frontside buried channel to collect signal holes is formed with a gate insulator layer, placed on top of a layer of p-doped material, itself located atop the n-type (bulk) substrate. The channel p-doping density and the  gate voltages are  chosen so that for a depleted device, a potential minimum (the ``buried channel'') is formed near the p-n interface due to the interplay of the potential bias and opposing electric fields sourced by the negative ions in the p-Si, as seen in Fig. \ref{fig:burried_channel} (left panel). Signal holes are collected in this minimum.  

\begin{figure}[ht]
\includegraphics[width=\columnwidth]{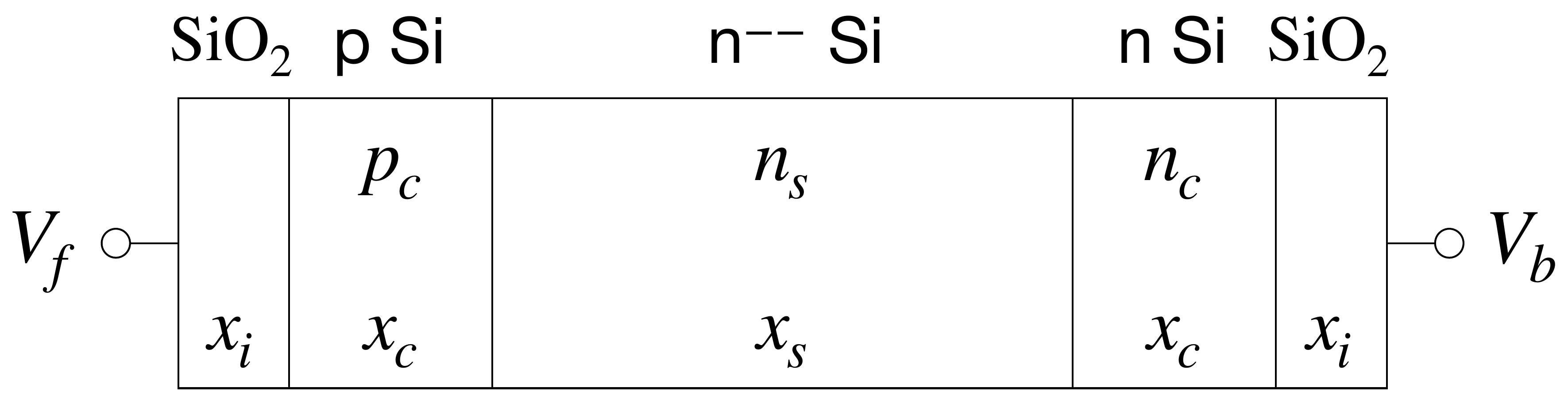}
\caption{Layers of a DCCD, from the front (left) to the back (right) of a pixel. $V_f$ and $V_b$ are front and backside gate voltages. 
The insulating layer, channel, and substrate widths are $x_i$, $x_c$, and $x_s$. The p and n-type doping concentrations of the channels and substrate are $p_c$, $n_c$, and~$n_s$.}
\label{fig:schematics}
\end{figure}

To equip the device with a  buried channel on the substrate's backside, we add a backside gate, an insulator layer, and a layer of n-Si with a doping density and thickness commensurate with the frontside's p-Si, as shown in Fig. \ref{fig:schematics}. The addition of the n-Si on the backside allows for the formation of the backside buried channel; under full depletion and for a negative front bias relative to the back gates, the positive ions of the n-Si induce an electric field that opposes the one set by the bias, creating a potential \textit{maximum}  where signal \textit{electrons} are collected, as in Fig. \ref{fig:burried_channel} (right panel). Intuitively, the addition of the backside's n-Si mirrors the p-type structure on the front, but with an opposite polarity. 

\subsection{Electrostatic Potential Profile}

To demonstrate the formation of the buried channels, we examine the idealized situation where the dopant concentrations are spatially constant within each layer.\footnote{We have checked with realistic dopant implantation simulations that this idealized calculation is sufficient to discuss the main characteristics of the dual channels.} Fig.~\ref{fig:schematics} shows five material layers that from left to right correspond to: front-side oxide, front-side p-Si, bulk $n^{--}$ Si, backside n-Si, and backside oxide. We label these layers with the letters A, B, C, D, and E, respectively, and solve for the electric fields across them. We work in a one-dimensional approximation along the depth of the DCCD. Using Gauss's law and continuity of the normal component of the  displacement field across dielectric interfaces, we obtain the electric fields as a function of the depth coordinate $x$ (with $x=0$ at the DCCD frontside),
\begin{eqnarray}
\nonumber E_A(x) & = &\Esio \\
\nonumber E_B(x) &=&\frac{\esio}{\esi} \Esio - \frac{qp_c (x-x_i)}{\esi}\\
\nonumber E_C(x) &=&\frac{\esio}{\esi} \Esio  - \frac{qp_c x_c}{\esi} + 
\frac{qn_s}{\esi}(x - x_c-x_i) \\
\nonumber E_D(x) &=&\frac{\esio}{\esi} \Esio  - \frac{qp_cx_c}{\esi} + 
\frac{qn_sx_s}{\esi} \\
\nonumber &&
+ \frac{q n_c} {\esi}(x-x_c-x_s-x_i) \\
 E_E(x) &=& \frac{\esi}{\esio} 
E_D(x_i+2x_c+x_s)\ ,
\label{eq:Esol}
\end{eqnarray}
where $q=\sqrt{4\pi \alpha}$ is the electron charge ($\alpha$ being the fine-structure constant). For the static dielectric constants we use the room-temperature values, $\esi=11.6$ and $\esio=4.4$~\cite{palik1998handbook}. $\Esio$ is the (constant) electric field on the front-side oxide. These equations must be complemented with the bias voltage boundary condition\footnote{In the presence of flat-band potentials these must be added or substracted from the front and backside gate voltages.}
\begin{equation}
\int_{0}^{2x_i+2x_c+x_s} dx \ E(x)= V_f-V_b \quad ,
\end{equation}
which can be used to solve for $\Esio$, resulting in
\begin{eqnarray}
\nonumber \Esio&=&
-\frac{1}{2 \esio (2 x_i \esi + (2x_c+ x_s)\esio)}
\\
\nonumber
&&\big[
q \esio ( n_c  x_c^2 
- 3  p_c  x_c^2  
+2  n_s  x_c x_s 
\\ \nonumber &&
- 2  p_c  x_c x_s
+  n_s  x_s^2 
) 
\\
\nonumber
&&
+q \esi
( 2 (n_c-p_c) x_c  x_i 
+ 2  n_s x_s x_i )
\\
&&
+
2 \esi \esio(V_b-V_f)
\big]\ .
\label{eq:sio2sol}
\end{eqnarray}
Using Eq.~\eqref{eq:sio2sol} in \eqref{eq:Esol} gives the solutions for the electric field and the corresponding electrostatic potential across the DCCD. 

\begin{figure}[ht]
\includegraphics[width=0.8\columnwidth]{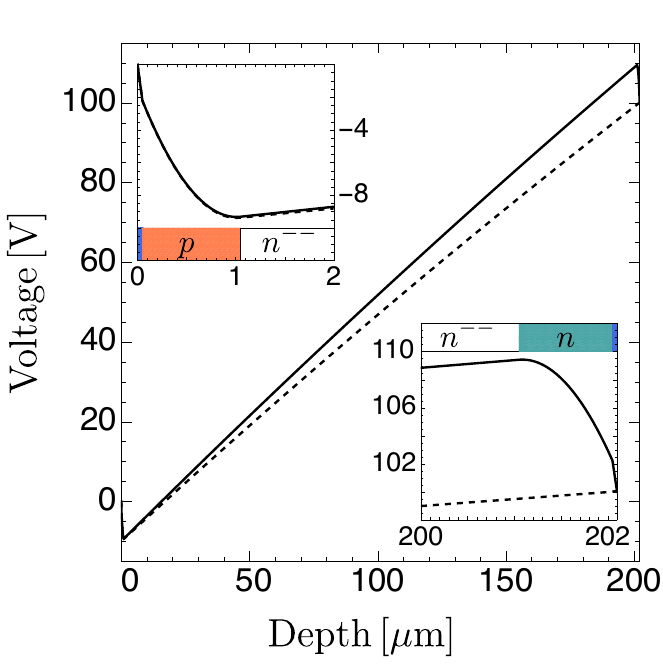}
\caption{\textit{Solid black}: Potential profile of a DCCD for $n_s=3\times 10^{11}\,\cm^{-3}$, $p_c=n_c=10^{16}\cm^{-3}$, $x_c=1\mum$, $x_i=50\nm$, $V_b=100$\,V and $V_f=0$\,V  (c.f. Fig. \ref{fig:schematics}). Front and backsides are located on the left and right of the plot. The upper-left and lower-right insets zoom in on the profiles around the potential extrema.  $p$, $n^{--}$, $n$ doped, and oxide regions are shown in orange, white, green, and blue. \textit{Dashed black}: potential of a standard p-channel-only CCD  obtained by removing the backside doping (setting  $n_c=n_s=3\times 10^{11} \cm^{-3}$) and oxide, shown for comparison.}
\label{fig:voltages}
\end{figure}

With the above solutions we present the potential profile of a DCCD design example by taking a high-resistivity substrate with thickness  $200\,\mum$ and n-doping density of $n_s=3\times 10^{11}\,\cm^{-3}$ (resistivity $\simeq 15\,\mathrm{k}\Omega\cdot\cm$), channel with thickness $x_c=1\mum$ and dopings $p_c=n_c=10^{16}\cm^{-3}$, and SiO$_2$ insulators with thickness $x_i=50\nm$. The bias between the backside and frontside is taken to be $V_b-V_f=100$\,V, while the reference frontside voltage is set to $V_f=0$\,V. The resulting potential profile is shown in   Fig.~\ref{fig:voltages}, where we see the emergence of a potential minimum and maximum on the front and backsides near the $p-n^{--}$ and  $n^{--}-n$ junctions, respectively, appropriately separated from the oxide interfaces. These distinct  extrema correspond to  the dual buried channels, which are the essential elements of our DCCD.  

Further insight into the solutions for the electrostatic potential can be obtained by taking advantage of the hierarchies between the substrate and channel or insulator thicknesses, $x_c/x_s, x_i/x_s \ll 1$, and between the substrate and channel doping concentrations $n_s/n_c,n_s/p_c\ll 1$, which allow us to find simple expressions for the channel  locations and depths by expanding on these small ratios. First, we find that with the above hierarchies and at leading order in the expansion parameters, the front and back channels are located at the $p-n^{--}$ and $n^{--}-n$ junctions, respectively, which is consistent with the results shown in Fig.~\ref{fig:voltages}, where we see that the potential extrema approximately align with the junctions. The depth of the potential of the front channel is given by the difference of the front gate voltage and the overall profile minimum $V_{\mathrm{min}}$, which at leading order is
\begin{equation}
V_f - V_{\mathrm{min}} \simeq q p_c \bigg(\frac{x_c^2}{2\epsilon_{\mathrm{Si}}}+\frac{x_c x_i}{\epsilon_{\mathrm{SiO}_2}}\bigg) \quad ,
\label{eq:approx1}
\end{equation}
where $\epsilon_{\mathrm{Si}}$, $\epsilon_{\mathrm{SiO}_2}$ are the permittivities of the substrate and oxide layers. The potential depth of the backside channel is similarly approximated by  
\begin{equation}
V_{\mathrm{max}}-V_b \simeq  
q n_c  \bigg(\frac{ x_c^2}{2 \epsilon_{\mathrm{Si}}}
+\frac{x_c x_i}{\epsilon_{\mathrm{SiO}_2}}
\bigg)
\quad ,
\label{eq:approx2}
\end{equation}
where $V_{\mathrm{max}}$ is the maximum of the potential. Taking the design parameters of Fig.~\ref{fig:voltages}, these expressions lead to well depths of $\approx 10$\,V, which approximately match the results presented in the Figure. Note that with the above approximations,  both the physical position and potential depth of the channels are independent of the bias potential $V_b-V_f$, and are set instead by the channel widths and doping concentrations, and insulator properties as in a standard fully depleted single-sided CCD \cite{holland2003fully}. 

Note also that within the approximations, the depths of the front (p-type) and backside (n-type) wells in our DCCD are equal to the corresponding p or n-type well depths of single-sided CCDs sharing the same p or n-type channel design parameters. This can be seen by noting that the DCCD's front  and backside well depths Eqns.~\eqref{eq:approx1} and \eqref{eq:approx2}, are independent of $n_c$ and $p_c$, respectively.  As an example, taking $n_c\rightarrow n_s$ our device reduces to a single-sided p-type CCD with a well depth that matches the frontside well depth of a  DCCD that has instead $n_c\gg n_s$. This can also be seen in Fig. \ref{fig:voltages} by comparing the well depths of the solid (DCCD) and dashed-black (single-sided p-type CCD) lines.  

For thin devices where substrate widths are of order $\sim \mathcal{O}(10) \mum$
 the approximations \eqref{eq:approx1}-\eqref{eq:approx2} are numerically inexact, but we have checked using the exact electrostatic solutions Eqns. \eqref{eq:Esol}-\eqref{eq:sio2sol}  that for such devices dual wells can still be easily designed with typical channel doping densities.

\subsection{Inversion}

As in a standard CCD, the DCCD can be inverted to reduce surface DCs \cite{saks1980technique,hynecek1981virtual} by populating the Si-oxide interfaces with minority carriers. We envision that the minority carriers on the interface between the oxide and n-Si on the backside could be provided by p-type channel stops, or if these are depleted, by other p-Si contacts. Similarly, the minority carriers at the oxide and p-Si interface on the frontside could stem from the n-type front channel stops or contacts.

\begin{figure}[ht!]
%\begin{adjustbox}{left=1\textwidth}
\includegraphics[width=0.8\columnwidth]{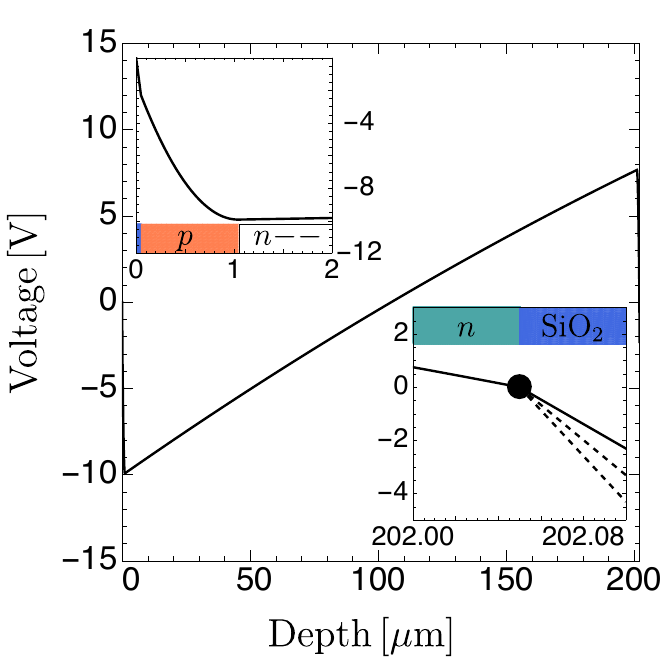}
%\end{adjustbox}
\caption{Potential profile of a DCCD (with a geometry and material coloring as in Fig.~\ref{fig:voltages}) with a frontside gate potential equal to $0\,$V and a backside gate potential that has been reduced to invert the backside channel. The upper-left inset zooms in the front channel potential, which is non-inverted, while the lower-right inset zooms in the n-Si/oxide interface at the inverted back channel. At that interface, the potential is pinned at $0$\,V, as shown by the black circle.
\label{fig:inversion1}}
\end{figure}

We illustrate the steps leading to inversion by considering a DCCD  with design parameters as in the previous section, namely $x_i=50\nm$, $x_c=1\mum$, $x_s=200\mum$, $n_c=p_c=10^{16}\cm^{-3}$, and $n_s=3\times 10^{11}\cm^{-3}$.  We assume that the minority carriers from the channel stops or extra contacts are kept at $0$~V. Starting from normal biasing conditions, as in Fig.~\ref{fig:voltages} ($V_f=0$\,V, $V_b=100$\,V), we begin by decreasing the backside voltage down to negative values. We find from our electrostatic solutions Eqns.~\eqref{eq:Esol}-\eqref{eq:sio2sol} that at $V_b= -2.3$\,V the potential at the n-Si/oxide interface is reduced to $0$\,V, as shown in Fig.~\ref{fig:inversion1}. At this point holes from the channel stops or contacts flow into the interface with the oxide, pinning the potential at that value. As in a standard CCD in inversion, further reductions of the backside gate potential are screened by the interface charges so that the bulk potential remains unaffected, as represented by the dashed black lines in  the figure, and the reductions are compensated by an increase in the magnitude of the oxide electric field. This leads to inversion of the backside channel. 

\begin{figure}[ht!]
%\begin{adjustbox}{left=1\textwidth}
\includegraphics[width=0.8\columnwidth]{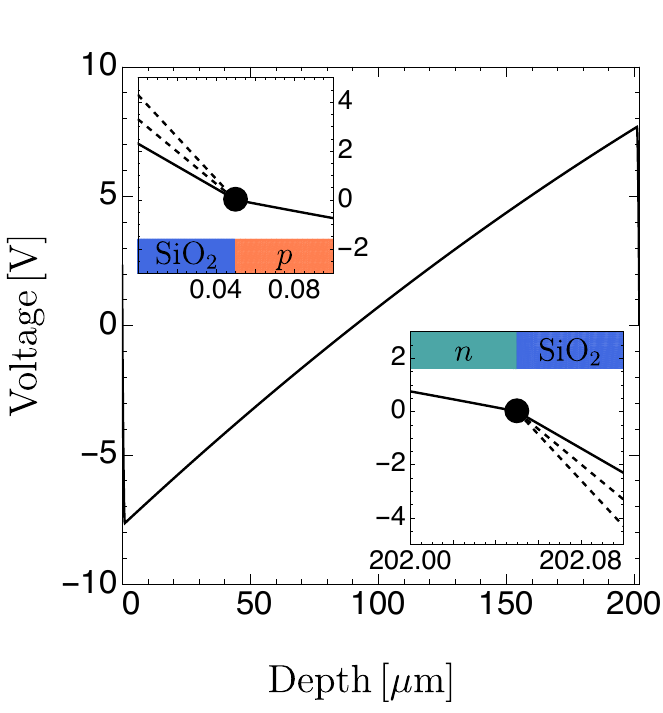}
%\end{adjustbox}
\caption{Potential profile of a DCCD (with a geometry as in Fig.~\ref{fig:voltages}) with front and backside gate potentials that have been respectively raised and lowered to invert both front and back channels. The insets at the upper-left and lower-right zoom into the doped-Si/oxide interfaces at both inverted channels, where the potentials are pinned at $0$\,V as shown by the black circles.
\label{fig:inversion2}}
\end{figure}

At this point the frontside channel can be inverted by raising the frontside gate potential. To calculate the point of inversion, we solve the first four electrostatic equations in Eq.~\eqref{eq:Esol}, which describe the electric field up to the interface with the backside oxide, subject to the boundary condition that the potential at the backside n-Si/oxide interface is pinned to $0$\,V,
\begin{equation}
\int_{0}^{x_i+2x_c+x_s} dx\ E(x)= V_f-0
\quad .
\end{equation}
By solving these equations, we find that at $V_f = 2.3$\,V, the potential at the frontside p-Si/oxide interface reaches $0$\,V. Here the n-type minority carriers flow into the interface and screen further reductions in the frontside potential, as shown in Fig.~\ref{fig:inversion2}. Note that the front and backside gate potentials required for front and back inversion are equal in magnitude but opposite in sign $V_f=-V_b=2.3$\,V, as expected from our symmetric design.  

A DCCD offers the possiblity of inverting both front and back channels, only the front channel, or only the back channel. This versatile operational capability could be valuable for the characterization of surface dark currents in CCDs. 

Current CCDs with record-low dark counts are operated by first driving the device into inversion to fill interface traps \cite{69907,SENSEI:2020dpa}. The device is then driven back to the non-inverted state to benefit from the efficient channel charge transfer and reduced lateral charge diffusion in this mode \cite{janesick1987scientific,holland2003fully}. Data is taken in non-inverted mode, but while interface traps are still filled. This procedure allows to exploit the main advantages of both inverted and non-inverted operation, and is critical to reduce dark counts and \textit{e.g.} increase the sensitivity of the device for dark-matter detection. We envision that under normal operating conditions a DCCD would function in an analogous way; first, the DCCD would be driven to dual-channel inversion to fill front and backside interface traps, and then data will be taken in non-inverted mode to facilitate charge transfer in both channels and to reduce charge diffusion.

\subsection{Readout}
\begin{figure}[t]
\centering
\includegraphics[width=0.99\columnwidth]{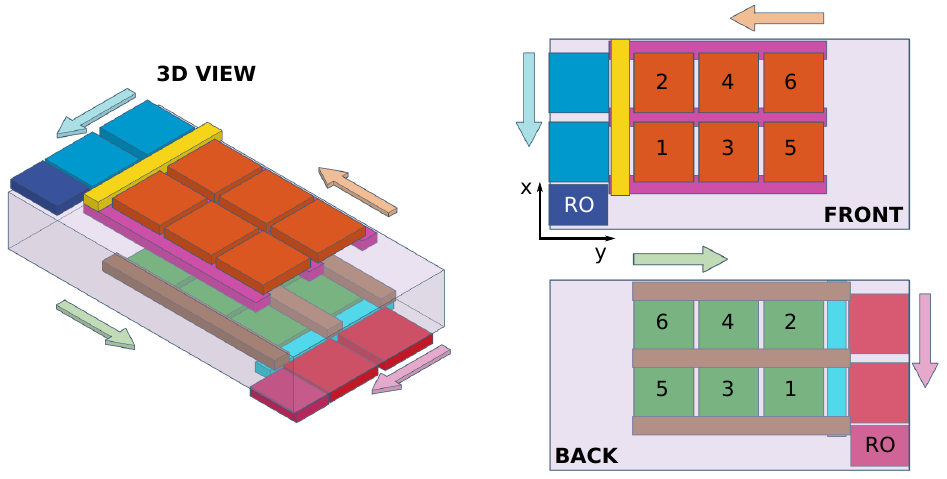} \caption{View of a 3x2 pixels DCCD. 
Front and backside pixels are shown in dark orange and green, channel stops appear in magenta and brown, and gates are shown in blue and red, respectively. Arrows indicate the direction of charge transfer towards readout.
The numbers indicate the pixel readout order on each side. The $x$ is the coordinate along both SRs, and $y$ along the VR of the front CCD.
}
\label{fig:basic_pix_structure}
\end{figure}

\begin{figure*}[ht!]
\centering
\includegraphics[width=0.9\textwidth]{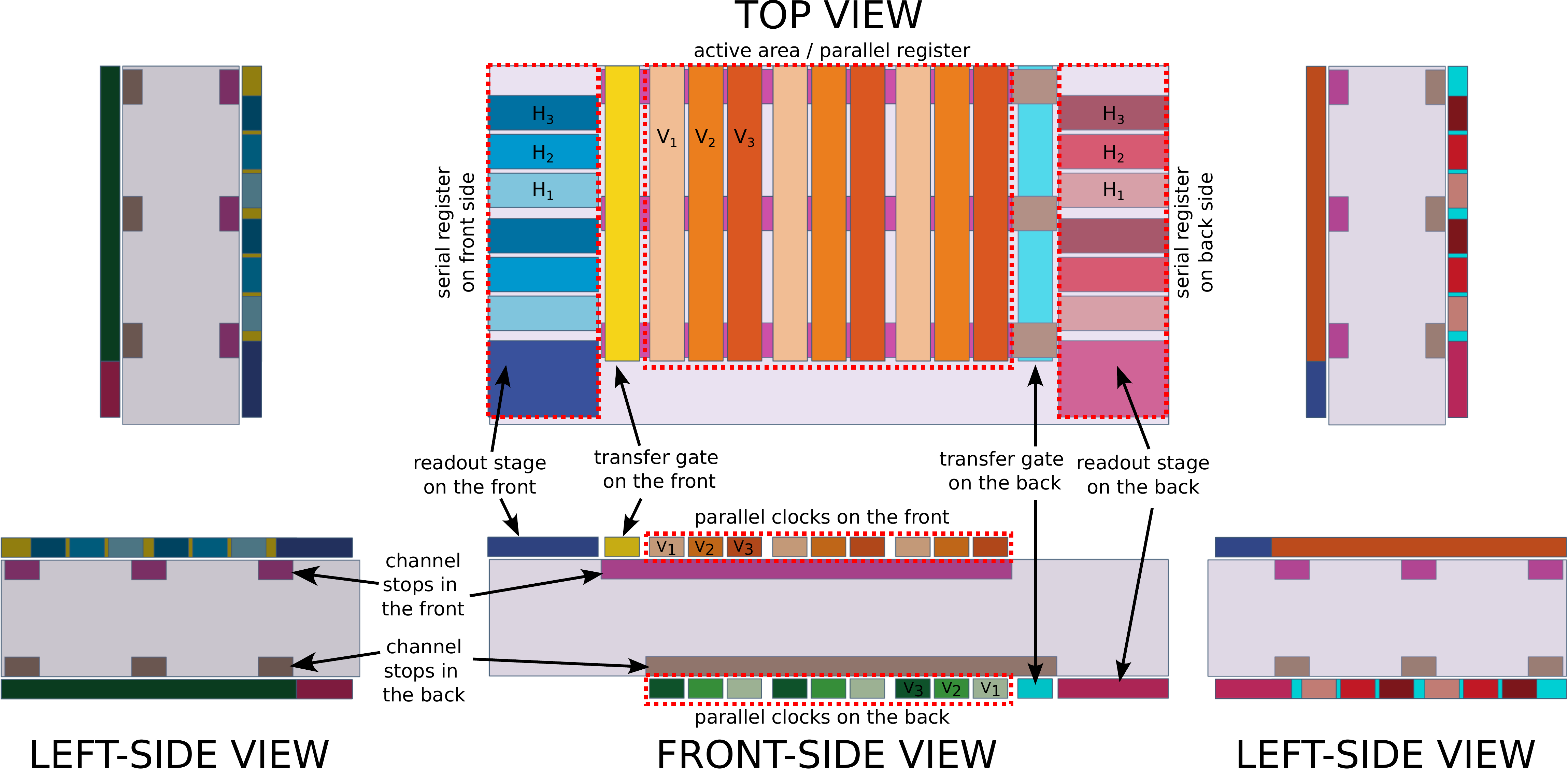}
\caption{Schematic view of a 3x2 pixels DCCD with three-phase CCDs on each side. The parallel clocks on the front side are shown in shades of orange and the ones in the back side are in shades of green (as in all the other pictures in this article). The three-phase gate voltages facilitating charge transfer down the vertical registers are labeled as $V_i$, $i=1,2,3$, while the three-phase voltages used to transfer charge along the SR are labelled as $H_i$, $i=1,2,3$. \label{fig:ckls_detail}
}
\end{figure*}

\begin{figure*}[ht!]
\centering
\includegraphics[width=0.7\textwidth]{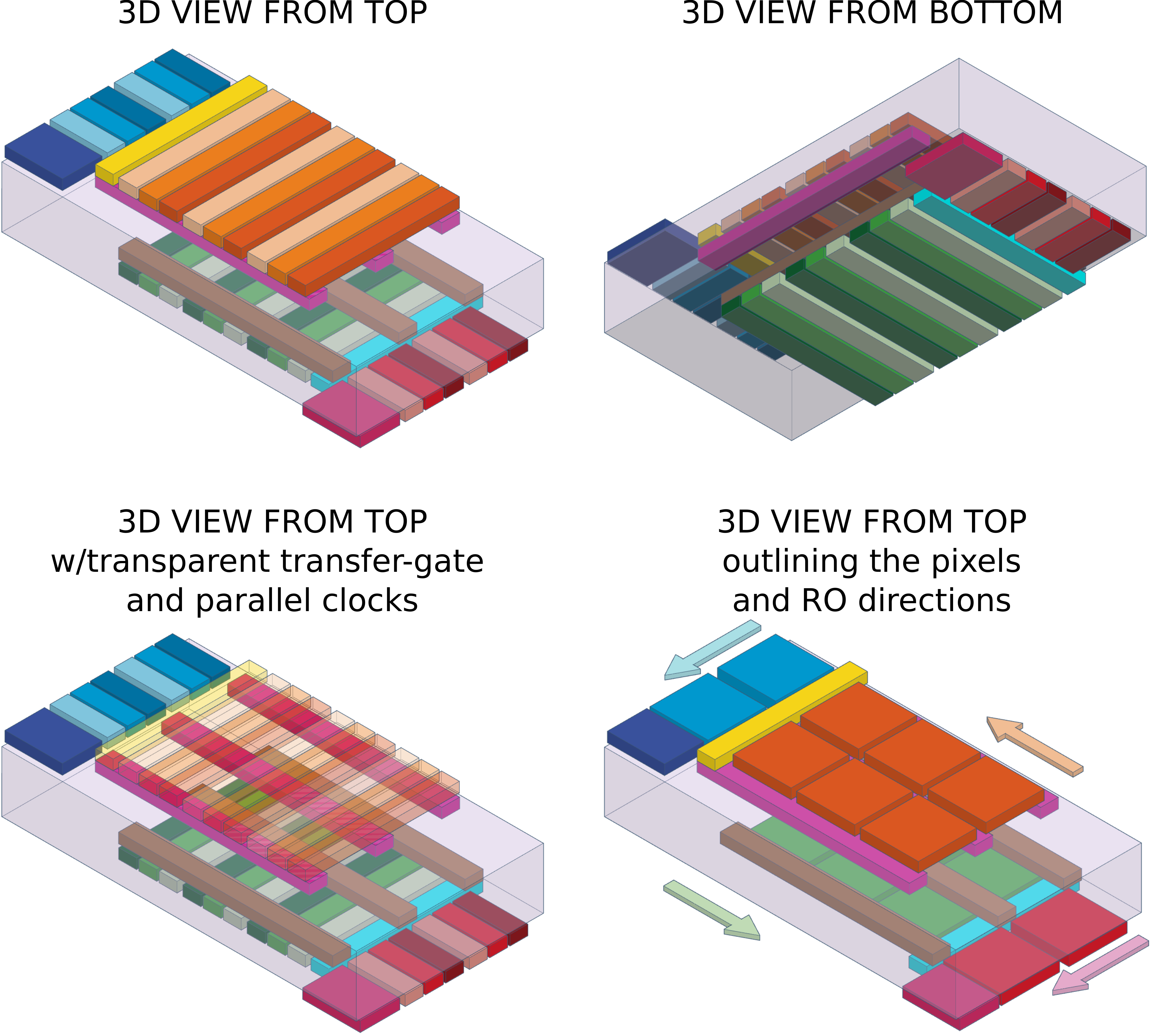}
\caption{\textit{Top two images:} Schematic 3D view of a 3x2 pixels DCCD with three-phase CCDs on each side. \textit{Bottom left:} Same as top-left but with transparent front transfer gate and parallel clocks to show internal structure. \textit{Bottom right:} Schematic view with the pixels outlined. Arrows indicate the direction of charge transfer leading to readout.}
\label{fig:ckls_3d}
\end{figure*}

Charge stored in the  channels is read out as in a standard CCD~\cite{janesick1987scientific}, so we only briefly comment on the additional design features that enable dual-charge readout.
In a DCCD there are both front and backside gates with synchronized clocked voltages that move the charges along front and back vertical (``parallel'') registers (VR) into two distinct serial registers (SR), one designed to read holes at the front, and the other dedicated for electrons at the back. Charge is confined within the VR by separate sets of channel stops at the front and back. A schematic design of the device is shown in Fig.~\ref{fig:basic_pix_structure} (left). In the DCCD, electrons and holes are shifted in \textit{opposite directions} along the VR; this feature enables enhanced timing as discussed below. With this scheme, the order in which charges are read is illustrated in Fig.~\ref{fig:basic_pix_structure} (right panel). Additional schematics are presented in Figs.~\ref{fig:ckls_detail} and ~\ref{fig:ckls_3d}, where we show transverse views along the different cross-sectional directions of the device and a three-dimensional representation of the device, respectively.

\section{Device Capabilities}

The DCCD can be operated in timed exposure or continuous readout mode, with each mode presenting  advantages that can be selected based on the specific application. 

\subsection{Timed Exposure Mode: \\ DC Rejection and Other Capabilities}

DCCDs distinguish surface or spurious charge backgrounds from bulk signals by discriminating single versus dual-channel events. Such backgrounds are thus reduced to the rare cases where front and back DCs coincide at the same sensor location within the exposure time. 

The single-channel DC rejection factor can be computed analytically under the approximation of large-CCD (negligible edge effects) and low occupancy (negligible DC overlap). Given a single-hole event in the front face, a coincident event is obtained if a corresponding single-electron event is registered in the back \footnote{Note that a multiple-electron event in the backside given a single hole in the front does not correspond to a coincident DC, as the number of charges on the front and back must match for the DCCD to mistake the event for a true ionization signal.}. This occurs with a probability $P_{n=1}=\lambda \exp(-\lambda)$. The non-coincidence probability is then given by 
\begin{eqnarray}
\nonumber P_{\textrm{non-coincidence}}&=&(1-P_{n=1})^{n_{\mathrm{pix}}}=[1-\lambda \exp(-\lambda)]^{n_{\mathrm{pix}}}\\ &\approx& 1-n_{\mathrm{pix}} \lambda 
\quad ,
\end{eqnarray}
where $\lambda$ is the expected number of single-channel DCs per pixel per exposure time, $n_{\mathrm{pix}}$ is the number of pixels around the frontside event that will be considered as a possible match in the backside, and the approximation is valid in the low-DC limit where $n_{\mathrm{pix}} \lambda\ll 1$. The parameter $n_{\mathrm{pix}}$ is introduced to account for the possibility that carriers created in a bulk event can move to adjacent pixels. The matching technique for these true events has to account for diffusion by enlarging the search ``radius'' $r$ around the collected carriers to find their complement on the other channel.  The diffusion radius $r$ defines a square centered on the event's frontside pixel location, with side-length equal to $2r+1$.  For thin sensors diffusion is negligible and $r=0$, but thick sensors can be several hundreds of microns across in which case diffusion can lead to inter-pixel migration. Following the results of \cite{SENSEI:2020dpa}, for sensors thicker than $\approx 600\mum$, $r=2$ should typically be used. For timed exposure (TE) operation the range of values that $n_{\mathrm{pix}}$ is thus given by
\begin{equation}
n_{\mathrm{pix}}^{\mathrm{TE}}=\left\{ \begin{array}{cc}
 1 & \textrm{no diffusion (thin sensor)}~r=0\\
9 & \textrm{1-pixel diffusion radius}~r=1\\
25 & \textrm{2-pixel diffusion radius}~r=2
\label{eq:TE}
\end{array} \right.
 \quad .
\end{equation}

The coincidence probability sets the DC rejection factor, and is given by 
\begin{eqnarray}
\nonumber P_{\textrm{coincidence}}&=&1-P_{\textrm{non-coincidence}} \label{eq:DCrejection} \\
 \nonumber 
 &=&1-[1-\lambda \exp(-\lambda)]^{n_{\mathrm{pix}}} \approx n_{\mathrm{pix}} \lambda \quad , \\
\end{eqnarray}
where again the approximation is valid for $n_{\mathrm{pix}} \lambda\ll 1$.

\begin{figure}[t]
\centering
\includegraphics[width=1\columnwidth]{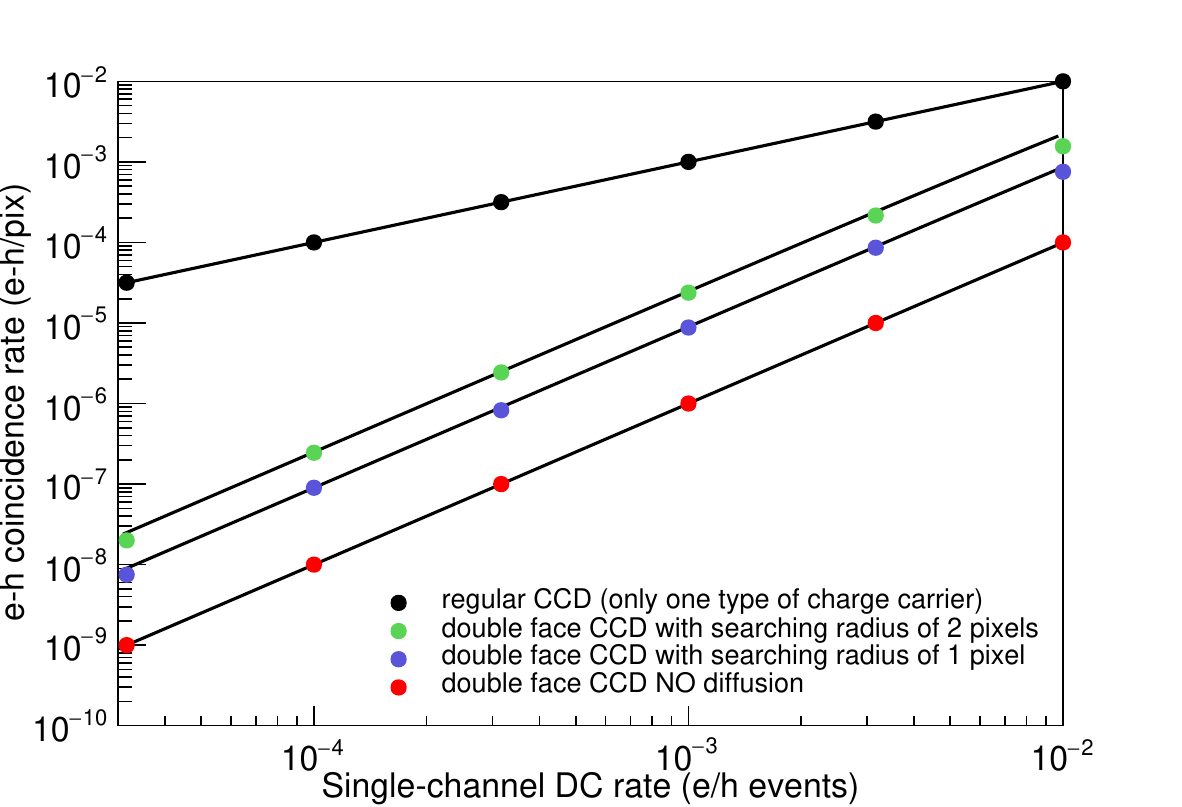}
\caption{Rate of coincident $e-h$ events versus the per-pixel single-channel DC rate within an exposure time ($\lambda$ in the text), assumed to be the same on both detector sides. Colored circles show the rates for different assumptions regarding bulk carrier diffusion; red corresponds to a search radius $r=0$, blue is $r=1$ and green is $r=2$ pixels. Black circles show the DC rates for a standard CCD. The plotted range of single-channel DCs represents realistic rates for day-long exposures \cite{SENSEI:2020dpa}. The solid black lines are  analytic calculations assuming an infinite CCD, Eq.~\eqref{eq:DCrejection}. Note that in a DCCD the DC rates depend \textit{quadratically} on the single-channel DC rate, since DCs are obtained only when front and back  DCs coincide in the same sensor location.}
\label{fig:single_e_event_reduction}
\end{figure}

To confirm our analytic results with a more realistic calculation that accounts for edge effects and DC overlaps, we perform single-$e/h$ DC Monte Carlo simulations on a  $1000 \times 1000$-pixel DCCD. The results are shown in Fig. \ref{fig:single_e_event_reduction}, where we see the rate of dual-channel DCs from the coincidences of single-channel events as a function of their per-pixel rate within an exposure time $\lambda$, assumed to be the same on both sensor sides. The dual-channel  rates are shown for different assumptions regarding the diffusion of bulk $e$ and $h$'s, \textit{i.e.}, for different search radii $r$.
The rates shown in the figure are thus the coincidences that occur within the search radius for each assumption. For each simulated point, the ratio between the horizontal and vertical coordinates represents the DCCD's single-channel DC rejection factor.

As expected, we find that the dual-channel rate for all diffusion scenarios is significantly lower than the single-channel one, with a larger reduction for lower single-channel DCs, since this leads to a lower probability of coincidental occurrences. In the absence of diffusion the dual-channel rate is simply the per-pixel single-channel rate squared, as expected from coincidences within a pixel. 
When diffusion is added, the dual-channel DCs increase, since the allowed area for coincidences is larger,  but we still find strong DC rejection. In all cases, we find that the coincidence DC rates obtained from the Montecarlo analysis are accurately described by  the analytic estimate in Eq.~\ref{eq:DCrejection}. Note that dual-channel surface DCs grow \textit{quadratically} with exposure time, as longer exposures increase the coincidence probabilities. For DCs of order $\lambda=10^{-4}$/pixel/day and day-long exposures as in~\cite{SENSEI:2020dpa}, our results show that a DCCD could lead to a three-order of magnitude reduction of single-channel DCs.  

Besides DC rejection, the DCCD can distinguish bulk events that occur during readout from those that occur during the exposure, as the former are registered as electrons or holes without a counterpart of opposite polarity within the same pixel. Charge-transfer inefficiencies are also single-channel issues, and can be mitigated using our dual-imaging strategy.

\subsection{Continuous Readout Mode: \\Timing and DC Rejection}

\begin{figure}[t!]
\centering
\includegraphics[width=0.9\columnwidth]{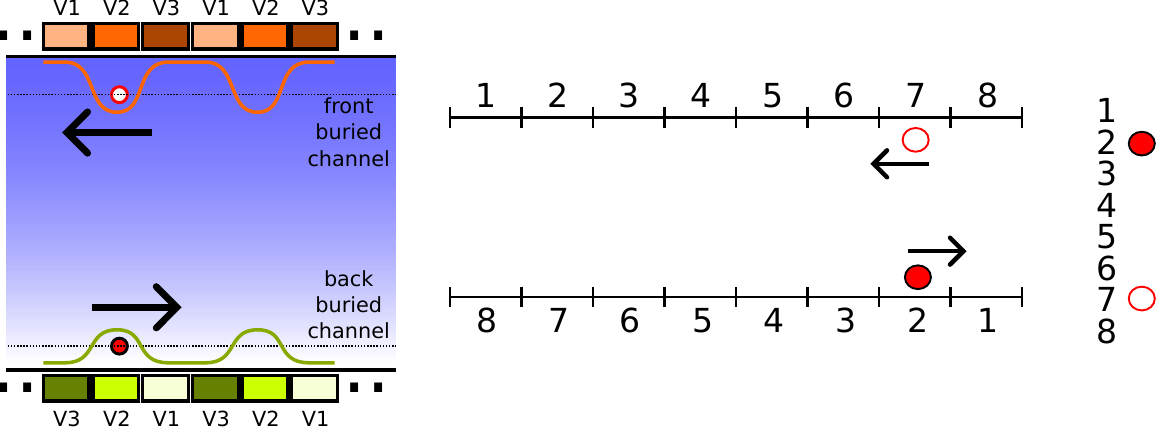}\hspace{0.5cm}
\caption{\textit{Left:} Cross-section of a DCCD, with electrons and holes, in the front and back VRs respectively, being moved in opposite directions towards readout. $V_{i},i=1,2,3$ are the three-phase voltages of a pixel. \textit{Right:} In continuous readout, one can determine the event vertical-register position by matching the back and front charges. There is only one position that produces the recorded values (2 and 7 in this example).}
\label{fig:dual_read}
\end{figure}

When standard CCDs are read out continuously an ambiguity between the location of the interaction on the VR and its time arises, as events that occur on a pixel at a given time can be mimicked by events that occurred earlier upstream in the VR, leading to a loss of vertical localization~\cite{de2022search,Crisler:2018gci}. DCCDs overcome this limitation by reading the front and back active areas in opposite directions. This allows one to disambiguate the absolute position of the primary interaction by looking for correlations in the recorded charge packages (up to event pileup, discussed below), as shown in Fig.~\ref{fig:dual_read}. 

For an event occurring at an absolute coordinate $y$ in the CCD parallel direction ({distance $y$ in pixels between the VR pixel location and the frontside SR}) and time $t$ from the start of the readout, the times of arrival $t_f$ ($t_b$) of the corresponding holes (electrons) at the front (back) SR are given by
\begin{eqnarray}
\nonumber t_f &=& y + t \\
t_b &=& L_{\mathrm{VR}}-y+ t
\quad ,
\end{eqnarray}
\noindent 
where $L_{\mathrm{VR}}$ is the physical number of pixels of the CCD in the vertical or parallel direction (the number of pixels in the VR) and times are measured in integer numbers of VR pixel shifts required to place charges in the SR. As charges stay on a VR pixel for the time it takes to read out the SR, $T_{\mathrm{SR}}$, times are thus quantified in units of $T_{\mathrm{SR}}$. 
Given the measured $t_f$ and $t_b$, from these equations we can reconstruct the event's position as
\begin{equation}
y = \frac{1}{2}(t_f - t_b + L_{\mathrm{VR}}) \quad .
\label{eq:position}
\end{equation}
Furthermore, from the event's position and the time of arrival at the SR the event's timestamp is determined by
\begin{equation}
t = t_f-y = \frac{1}{2}(t_f+t_b-L_{\mathrm{VR}}) \quad ,
\label{eq:time}
\end{equation}
with a time resolution $T_{\mathrm{SR}}$ \footnote{We neglect the $e-h$ drift times to the channels, which are typically 5 orders of magnitude shorter than the clocking times, see e.g. \cite{Du:2020ldo}}. Thus, while an event's timing in a standard CCD is given by the time it takes to read the whole VR, $T_{\mathrm{CCD}}$, in our DCCD this is \textit{significantly improved} as $T_{\mathrm{SR}}\approx  T_{\mathrm{CCD}}/L_{\mathrm{VR}}\ll T_{\mathrm{CCD}}$, especially in large devices where $L_{\mathrm{VR}}\gg 1$. In megapixel arrays $L_{\mathrm{VR}}\simeq 10^3$ and this represents a three-order of magnitude improvement with respect to conventional devices. 

Correlations between electron and holes, however, are only unique up to event pileup. 
Pileup occurs when the time difference between the events matches the time it takes to move their charges between their VR locations, in which case their charge packets are summed. Either the $h$ \textit{or} $e$ packets of different events can be summed, leading to overlapping events in the front \textit{or} backside images, respectively. This issue can be mitigated by correlating the front- and backside images.\footnote{If both front and back images overlap the events are classified as a single combined one. In particular, exposure to continuous sources such as stars leads to loss of vertical localization, since these sources continuously produce events in the VR pixels as the charges are being transferred, leading to front and back overlap with other events in those registers.} 
A second type of pileup arises when events with the same number of $e$ or $h$'s occur in a VR within $T_{\mathrm{CCD}}$, in which case 
multiple pairings between the $e$ and $h$ packets may be possible and the event's location and time are lost in pairing ambiguities (but it is still possible to veto such events). Optical/near IR photons always lead to single $e-h$ pairs, so pairing ambiguities often arise if another such photon or DC falls in a given VR within $T_{\mathrm{CCD}}$. 
X-ray events, on the other hand, rarely lead to the same number of $e-h$ pairs and are thus less prone to pairing ambiguities. 

\begin{figure}[t]
\centering
\includegraphics[width=0.99\columnwidth]{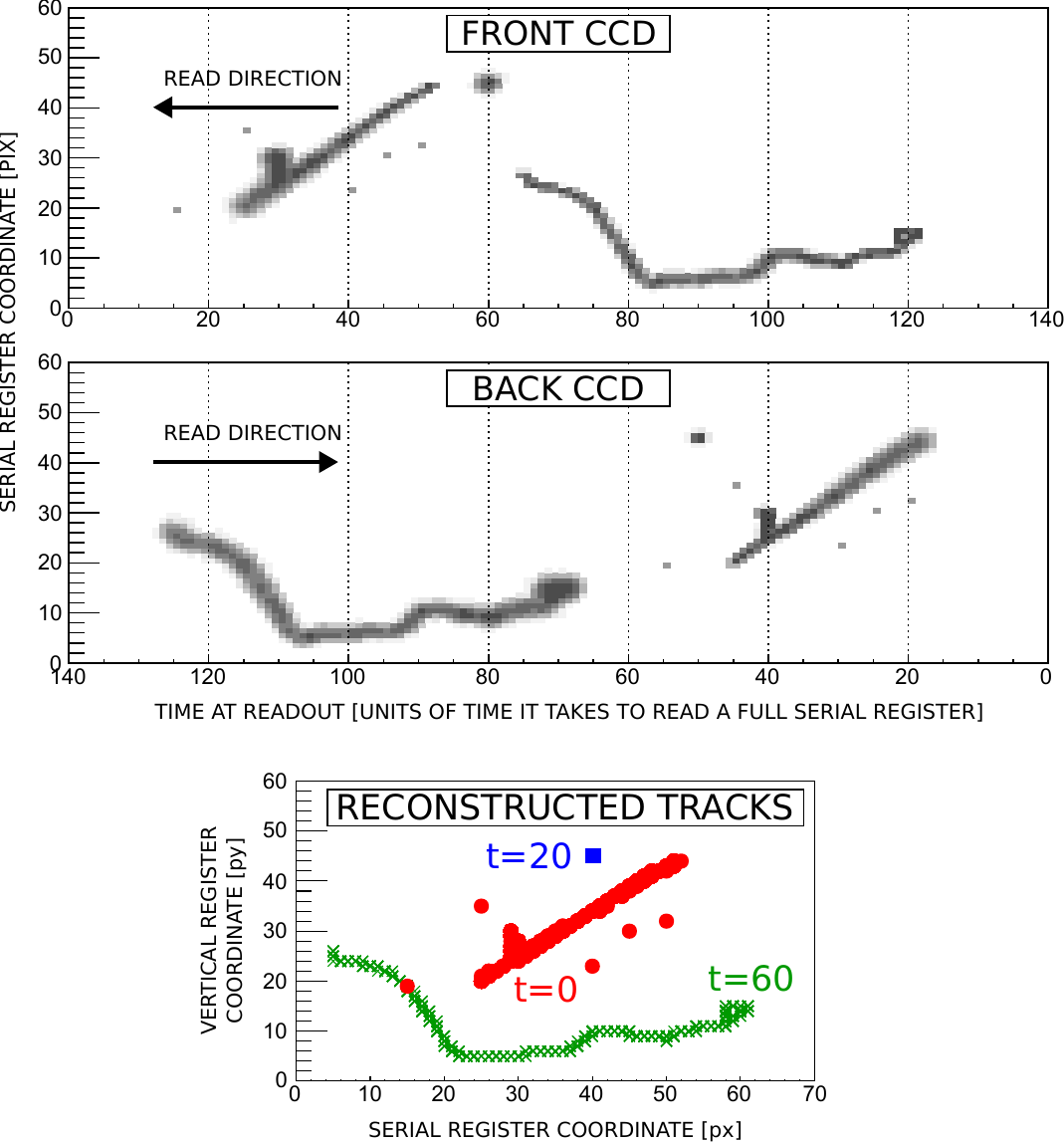}
\caption{Simulated events as seen by the front (top panel) and back (middle panel) readout stages of a DCCD. Serial and vertical registers are along the vertical and horizontal directions, respectively. Note that the diffusion of the charge carriers is mirrored. The lower panel shows the reconstructed image. The colors indicate the time of arrival in units of $T_{\mathrm{SR}}$.}
\label{fig:sim}
\end{figure}

The above considerations indicate that our DCCD's timing is most effectively utilized in situations with low illuminations, where pileup is limited. To exemplify a concrete application, we present in Fig.~\ref{fig:sim} a GEANT4~\cite{GEANT4:2002zbu} simulation in a $70\times 60$ pixel DCCD of three energetic events, one being spot-shaped, and two being track-like, with the shortest one  leading to isolated single $e-h$ events by \textit{e.g.}~secondary luminescence ~\cite{Du:2020ldo}. Given $L_{\mathrm{VR}}=70$, for the lower-left end of the shorter track we register $t_f=25$ and $t_b=45$, which using Eqns.~\eqref{eq:position} and \eqref{eq:time} results in  $y=25$ pixels and $t=0$ in units of $T_{\mathrm{SR}}$. Similar arguments indicate that the secondary photons happened concurrently, from which they can be associated to the track, while the spot-shaped X-ray and the longer track occurred at $t = 20 \, T_{\mathrm{SR}}$ and $t = 60 \, T_{\mathrm{SR}}$, respectively. The spot-like event in the image recorded by the front CCD would be considered as part of the halo of the high energy event in a standard device but the timing provided by the DCCD allows it to be separated.

In continuous readout, single-channel DCs can mimic signal events when front and back DCs coincide within $T_{\mathrm{CCD}}$ in a given VR. Since these coincidences are more likely than the per-pixel or search radius ones leading to DCs in timed exposure operation, (and given that continuous readout results in increased spurious charge~\cite{Crisler:2018gci,SENSEI:2020dpa}), in this mode the DC rejection capabilities are partially reduced when compared to timed exposure operation. In other words, the number of pixels $n_{\mathrm{pix}}$ that need to be considered for a front-back single-channel DC match (c.f. Eq. \eqref{eq:DCrejection}) is \textit{larger} in continuous readout mode than in timed exposure operation, and this results in a larger coincidence DC rate. Accounting for the coincidences within a whole VR, in continuous readout (CR) we have
\begin{equation}
n_{\mathrm{pix}}^{\mathrm{CR}}=\left\{ \begin{array}{cc}
 L_{\mathrm{VR}} & \textrm{no diffusion (thin sensor)}~r=0\\
3L_{\mathrm{VR}} & \textrm{1-pixel diffusion radius} ~r=1\\
5L_{\mathrm{VR}} & \textrm{2-pixel diffusion radius} ~r=2
\label{eq:CR}
\end{array} \right. \quad ,
\end{equation} 
where $L_{\mathrm{VR}}$ is the VR length in pixels. Using Eqns.~\eqref{eq:TE} and~\eqref{eq:CR} in Eq.~\eqref{eq:DCrejection}, we see that in the low-DC limit $n_{\mathrm{pix}} \lambda\ll 1$, single-channel DC rejection in timed exposure mode is stronger than in continuous readout by a factor $\sim L_{\mathrm{VR}}$ ($ \sim 10^3$ in megapixel arrays).
Nonetheless, sizable DC rejection capabilties can still be achieved in continuous readout operation;
for typical values $\lambda=10^{-4}$/pixel/day with day-long exposures and $ L_{\mathrm{VR}}\sim 10^{3}$ as in~\cite{SENSEI:2020dpa}, one order-of magnitude single-channel DC suppression can be achieved in thin sensors under continous readout.

\section{Dark Matter Reach}

In this section, we compare the anticipated dark matter (DM) detection capabilities of forthcoming large-exposure Skipper CCD-based experiments, like \cite{aguilar2022oscura}, with those of equivalent experiments that use instead Dual-Sided Skipper CCDs (Skipper DCCDs). To perform projections we follow the procedure in \cite{Du:2023soy}, briefly summarized here as follows. We consider experiments looking for DM scattering on electrons by measuring ionization signals in CCD pixels. We focus on models where DM scatters with electrons via a light mediator, \textit{i.e.}, models with a scattering form factor $F_{\textrm{DM}}(q)=1/q^2$ where $q$ is the momentum transfer \cite{Essig:2011nj}. We parametrize the strength of the scattering rate by the cross section on electrons $\bar{\sigma}_e$ at a fixed momentum transfer $q=\alpha m_e$, where $\alpha$ is the fine-structure constant and $m_e$ the electron mass \cite{Essig:2011nj}. For a given $\bar{\sigma}_e$ and DM mass, we compute the number of ionized electron-hole pairs in Si using QCDark \cite{Dreyer:2023ovn}. Based on the number of ionized electron-hole pairs for a DM signal event within pixels of the CCDs, we divide the search in one to four electron bins; the precise definition of each signal bin and the corresponding analysis cuts follows closely the strategy used by the SENSEI experiment at the MINOS cavern \cite{SENSEI:2020dpa}. 

We project limits when the rate of DM ionization events for any of the electronic bins exceeds the rate expected in the background-only hypothesis, accounting for statistical uncertainties at $95\%$ CL. The background hypothesis is obtained from the assessment of dark counts in Skipper CCD-based DM experiments presented in \cite{Du:2023soy}, and is summarized as follows. For the one-electron bin, following the analysis in \cite{SENSEI:2020dpa} we drop backgrounds coming from spurious charge, as these are measured and removed from the analysis. Consistent with the measured data in \cite{SENSEI:2020dpa} and the simulations performed in \cite{Du:2023soy}, we assume that the remaining one-electron rate has two components. First, it has a component due to low-energy Cherenkov photons with a rate $R_{1e}=185/$gram/day. We assume that forthcoming detectors will be able to remove radiative backgrounds with improved radiopurity and shielding (see \text{e.g.} \cite{aguilar2022oscura}), so we simply neglect this component. The second component is non-radiative in origin, is homogeneous across the whole CCD, and has a magnitude $R_{1e}=300/$gram/day. While its origin is unknown, as argued in \cite{Du:2023soy} and in this manuscript, it is likely associated with detector dark counts that occur on surfaces. Since this type of dark counts can be mitigated by the dual-readout strategy proposed in this manuscript, we assume that this background component is suppressed in DCCDs by a factor of $10^3$ with respect to its rate in standard (single-sided) Skipper CCDs, consistent with the estimates presented in Fig.~\ref{fig:single_e_event_reduction}. We acknowledge, however, that assuming that all DCs arise on the surface is a favorable assumption for our device, as it maximizes its DC rejection capabilities. 

For the higher electron bins, following the results in \cite{Du:2023soy} we assume that the 2, 3 and 4 electron background rates are dominated by coincidences of the aforementioned surface one-electron background events and spurious charge. As in the analysis of \cite{SENSEI:2020dpa} we do not perform any background substraction for these bins. This amounts to taking $R_{2e}=0.51$/gram/day,  $R_{3e}=0.035$/gram/day and $R_{4e}=4.22\times 10^{-6}$/gram/day in  single-sided Skipper CCDs, correspondingly \cite{Du:2023soy}. Since the one-electron dark count rate is dominated by surface events, and since spurious charge is associated with clock-voltage swings that also occur near detector surfaces \cite{janesick1987scientific}, under our assumptions all multi-electron events stem from coincidences of one-electron single-channel events. As a consequence, the DCCD is effective at suppressing multi-electron backgrounds, and the $10^3$ reduction of the single-electron backgrounds considered above translates into  a $10^{3n}$ reduction in the backgrounds for the $n$-electron bins $n=2,3,4$, with respect to the rates in single-sided Skipper CCDs.

\begin{figure}[ht!]
%\begin{adjustbox}{left=1\textwidth}
\includegraphics[width=1\columnwidth]{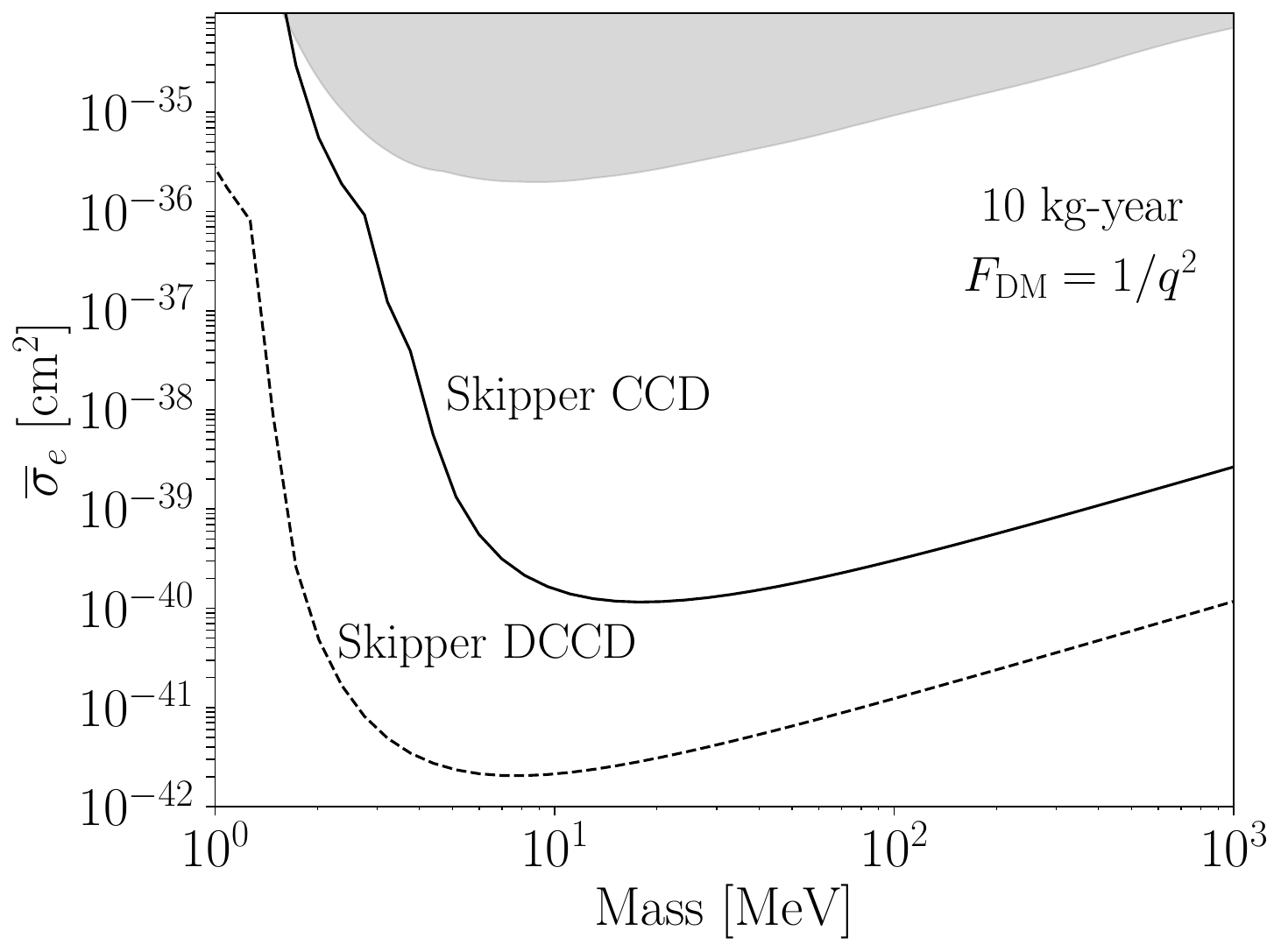}
%\end{adjustbox}
\caption{Projected exclusion regions at $95\%$ CL for a light-mediator DM model (scattering form factor $F_{\textrm{DM}}=1/q^2$) for an experiment using standard single-sided Skipper CCDs (solid black) and the Skipper DCCDs proposed in this manuscript (dashed black). The detector exposure is taken to be 10-kg-years, and backgrounds are accounted for as described in the body of the text. Gray shaded regions are already excluded by existing direct-detection experiments, with leading constraints coming from the Skipper CCD-based experiments \cite{SENSEI:2020dpa,SENSEI:2023zdf,DAMIC-M:2023hgj,DAMIC-M:2023gxo}. }
\label{fig:limitsDCCD}
\end{figure}

We perform projections assuming an exposure of 10-kg-years, which is the order-of-magnitude target exposure of future Skipper CCD-based detectors \cite{aguilar2022oscura}. Our results are presented in Fig. \ref{fig:limitsDCCD}. The figure shows that experiments using Skipper DCCDs have the potential to deliver significant improvements in the reach to detect DM with respect to those based on standard single-sided Skipper CCDs, due to the DCCD's enhanced background mitigation capabilities. 
 The most significant improvements are achieved for lower DM masses, as in these models the search is dominated by the low energy bins that have comparatively larger backgrounds than the higher-electron bins.  Quantitatively, we find that Skipper DCCDs could outperform the reach of standard Skipper-CCD by factors of $\approx 10^3$, $\approx 80$ and $\approx 20$ for DM masses of $1$, $10$ and $100$ MeV, correspondingly.

\section{Discussion}

We presented the basic design and capabilities of the ``Dual-sided CCD'' (DCCD), a device that collects both the electrons and holes created in an ionization event to deliver improvements in  DC rejection and timing with respect to  standard CCDs. By performing simulations of realistic applications, we showed that a DCCD is able to typically provide three-order of magnitude improvements in DCs and timing with respect to existing megapixel devices.

This article begins the study of dual-sided imagers, and opens up significant opportunities for novel research. First, the advancements brought by DCCDs have far-reaching implications for particle and astrophysics (see Introduction).  Second, the DCCD architecture can be integrated with several existing CCD designs, further enhancing its relevance. A DCCD can be fabricated on thick, high-resistivity Si to  improve the red response~\cite{holland1996200,stover1997characterization,holland2003fully}. A Skipper-DCCD can be designed to reduce readout noise ~\cite{janesick1990new,1050535,chandler1990sub,Tiffenberg:2017aac}. A dual-sided EMCCD~\cite{jerram2001llccd} could be fabricated to achieve sub-millisecond timing and would drastically reduce spurious charge~\cite{daigle2010darkest,daigle2012characterization,bush2021measurement}. Given the similarities between CCDs and CMOS imagers several of the design features discussed here may also be applied in the context of CMOS detectors to reduce surface DCs~\cite{10.1088/978-0-7503-3235-4}. 
Finally, this work motivates efforts to begin the fabrication of a DCCD, which can be manufactured based on existing and demonstrated designs for standard CCDs~\cite{aguilar2022oscura} by replicating these layouts on both sides of a Si wafer. A preliminary assessment of the fabrication timeline indicates that subject to funding, a first working prototype could be available within 3 years. 

\begin{acknowledgments}

The authors would like to thank Steve Holland for useful discussions, and Ezequiel Alvarez and the participants of the Voyages Beyond the SM III workshop for discussions during the early stages of this work. 
The work of P.D. is supported by the US Department of Energy under grant DE-SC0010008.
DEU is supported by Perimeter Institute for Theoretical Physics and
by the Simons Foundation. Research at Perimeter Institute is supported in part by the Government of Canada
through the Department of Innovation, Science and Economic Development Canada and by the Province of Ontario through the Ministry of Colleges and Universities.
R.E.~acknowledges support from DoE Grant DE-SC0009854, Simons Investigator in Physics Award 623940, and the US-Israel Binational Science Foundation Grant No.~2016153. 
R.E.~and J.T.~acknowledge support from the Heising-Simons Foundation under Grant No.~79921. This work was supported by Fermilab under U.S.~Department of Energy (DOE) Contract No.~DE-AC02-07CH11359. 

\end{acknowledgments}

\bibliography{CCD}

\providecommand{\href}[2]{#2}\begingroup\raggedright\begin{thebibliography}{10}

\bibitem{6768140}
W.~S. Boyle and G.~E. Smith, \emph{Charge coupled semiconductor devices},
  \href{https://doi.org/10.1002/j.1538-7305.1970.tb01790.x}{\emph{The Bell
  System Technical Journal} {\bfseries 49} (1970) 587--593}.

\bibitem{amelio1970experimental}
G.~F. Amelio, M.~F. Tompsett and G.~E. Smith, \emph{Experimental verification
  of the charge coupled device concept}, {\emph{Bell System Technical Journal}
  {\bfseries 49} (1970) 593--600}.

\bibitem{damerell1981charge}
C.~Damerell, F.~Farley, A.~Gillman and F.~Wickens, \emph{Charge-coupled devices
  for particle detection with high spatial resolution}, {\emph{Nuclear
  Instruments and Methods in Physics Research} {\bfseries 185} (1981) 33--42}.

\bibitem{janesick1987scientific}
J.~R. Janesick, T.~Elliott, S.~Collins, M.~M. Blouke and J.~Freeman,
  \emph{Scientific charge-coupled devices}, {\emph{Optical Engineering}
  {\bfseries 26} (1987) 692--714}.

\bibitem{Tiffenberg:2017aac}
{\scshape SENSEI} collaboration, J.~Tiffenberg, M.~Sofo-Haro, A.~Drlica-Wagner,
  R.~Essig, Y.~Guardincerri, S.~Holland et~al., \emph{{Single-electron and
  single-photon sensitivity with a silicon Skipper CCD}},
  \href{https://doi.org/10.1103/PhysRevLett.119.131802}{\emph{Phys. Rev. Lett.}
  {\bfseries 119} (2017) 131802},
  [\href{https://arxiv.org/abs/1706.00028}{{\ttfamily 1706.00028}}].

\bibitem{Crisler:2018gci}
{\scshape SENSEI} collaboration, M.~Crisler, R.~Essig, J.~Estrada,
  G.~Fernandez, J.~Tiffenberg, M.~Sofo~haro et~al., \emph{{SENSEI: First
  Direct-Detection Constraints on sub-GeV Dark Matter from a Surface Run}},
  \href{https://doi.org/10.1103/PhysRevLett.121.061803}{\emph{Phys. Rev. Lett.}
  {\bfseries 121} (2018) 061803},
  [\href{https://arxiv.org/abs/1804.00088}{{\ttfamily 1804.00088}}].

\bibitem{SENSEI:2019ibb}
{\scshape SENSEI} collaboration, O.~Abramoff et~al., \emph{{SENSEI:
  Direct-Detection Constraints on Sub-GeV Dark Matter from a Shallow
  Underground Run Using a Prototype Skipper-CCD}},
  \href{https://doi.org/10.1103/PhysRevLett.122.161801}{\emph{Phys. Rev. Lett.}
  {\bfseries 122} (2019) 161801},
  [\href{https://arxiv.org/abs/1901.10478}{{\ttfamily 1901.10478}}].

\bibitem{SENSEI:2020dpa}
{\scshape SENSEI} collaboration, L.~Barak et~al., \emph{{SENSEI:
  Direct-Detection Results on sub-GeV Dark Matter from a New Skipper-CCD}},
  \href{https://doi.org/10.1103/PhysRevLett.125.171802}{\emph{Phys. Rev. Lett.}
  {\bfseries 125} (2020) 171802},
  [\href{https://arxiv.org/abs/2004.11378}{{\ttfamily 2004.11378}}].

\bibitem{SENSEI:2021hcn}
{\scshape SENSEI} collaboration, L.~Barak et~al., \emph{{SENSEI:
  Characterization of Single-Electron Events Using a Skipper Charge-Coupled
  Device}}, \href{https://doi.org/10.1103/PhysRevApplied.17.014022}{\emph{Phys.
  Rev. Applied} {\bfseries 17} (2022) 014022},
  [\href{https://arxiv.org/abs/2106.08347}{{\ttfamily 2106.08347}}].

\bibitem{DAMIC:2016lrs}
{\scshape DAMIC} collaboration, A.~Aguilar-Arevalo et~al., \emph{{Search for
  low-mass WIMPs in a 0.6 kg day exposure of the DAMIC experiment at SNOLAB}},
  \href{https://doi.org/10.1103/PhysRevD.94.082006}{\emph{Phys. Rev. D}
  {\bfseries 94} (2016) 082006},
  [\href{https://arxiv.org/abs/1607.07410}{{\ttfamily 1607.07410}}].

\bibitem{Castello-Mor:2020jhd}
{\scshape DAMIC-M} collaboration, N.~Castell\'o-Mor, \emph{{DAMIC-M Experiment:
  Thick, Silicon CCDs to search for Light Dark Matter}},
  \href{https://doi.org/10.1016/j.nima.2019.162933}{\emph{Nucl. Instrum. Meth.
  A} {\bfseries 958} (2020) 162933},
  [\href{https://arxiv.org/abs/2001.01476}{{\ttfamily 2001.01476}}].

\bibitem{DAMIC:2020cut}
{\scshape DAMIC} collaboration, A.~Aguilar-Arevalo et~al., \emph{{Results on
  low-mass weakly interacting massive particles from a 11 kg-day target
  exposure of DAMIC at SNOLAB}},
  \href{https://doi.org/10.1103/PhysRevLett.125.241803}{\emph{Phys. Rev. Lett.}
  {\bfseries 125} (2020) 241803},
  [\href{https://arxiv.org/abs/2007.15622}{{\ttfamily 2007.15622}}].

\bibitem{DAMIC-M:2023gxo}
{\scshape DAMIC-M} collaboration, I.~Arnquist et~al., \emph{{First Constraints
  from DAMIC-M on Sub-GeV Dark-Matter Particles Interacting with Electrons}},
  \href{https://doi.org/10.1103/PhysRevLett.130.171003}{\emph{Phys. Rev. Lett.}
  {\bfseries 130} (2023) 171003},
  [\href{https://arxiv.org/abs/2302.02372}{{\ttfamily 2302.02372}}].

\bibitem{Du:2020ldo}
P.~Du, D.~Egana-Ugrinovic, R.~Essig and M.~Sholapurkar, \emph{{Sources of
  Low-Energy Events in Low-Threshold Dark-Matter and Neutrino Detectors}},
  \href{https://doi.org/10.1103/PhysRevX.12.011009}{\emph{Phys. Rev. X}
  {\bfseries 12} (2022) 011009},
  [\href{https://arxiv.org/abs/2011.13939}{{\ttfamily 2011.13939}}].

\bibitem{Oscura:2022vmi}
{\scshape Oscura} collaboration, A.~Aguilar-Arevalo et~al., \emph{{The Oscura
  Experiment}},  \href{https://arxiv.org/abs/2202.10518}{{\ttfamily
  2202.10518}}.

\bibitem{Oscura:2023qik}
{\scshape Oscura} collaboration, B.~A. Cervantes-Vergara et~al.,
  \emph{{Skipper-CCD Sensors for the Oscura Experiment: Requirements and
  Preliminary Tests}},  \href{https://arxiv.org/abs/2304.04401}{{\ttfamily
  2304.04401}}.

\bibitem{Oscura:2023qch}
{\scshape Oscura} collaboration, S.~Perez et~al., \emph{{Early Science with the
  Oscura Integration Test}},
  \href{https://arxiv.org/abs/2304.08625}{{\ttfamily 2304.08625}}.

\bibitem{Du:2022dxf}
P.~Du, D.~Ega\~na Ugrinovic, R.~Essig and M.~Sholapurkar, \emph{{Doped
  Semiconductor Devices for sub-MeV Dark Matter Detection}},
  \href{https://arxiv.org/abs/2212.04504}{{\ttfamily 2212.04504}}.

\bibitem{CONNIE:2019swq}
{\scshape CONNIE} collaboration, A.~Aguilar-Arevalo et~al., \emph{{Exploring
  low-energy neutrino physics with the Coherent Neutrino Nucleus Interaction
  Experiment}}, \href{https://doi.org/10.1103/PhysRevD.100.092005}{\emph{Phys.
  Rev. D} {\bfseries 100} (2019) 092005},
  [\href{https://arxiv.org/abs/1906.02200}{{\ttfamily 1906.02200}}].

\bibitem{Fernandez-Moroni:2020yyl}
G.~Fernandez-Moroni, P.~A.~N. Machado, I.~Martinez-Soler, Y.~F. Perez-Gonzalez,
  D.~Rodrigues and S.~Rosauro-Alcaraz, \emph{{The physics potential of a
  reactor neutrino experiment with Skipper CCDs: Measuring the weak mixing
  angle}}, \href{https://doi.org/10.1007/JHEP03(2021)186}{\emph{JHEP}
  {\bfseries 03} (2021) 186},
  [\href{https://arxiv.org/abs/2009.10741}{{\ttfamily 2009.10741}}].

\bibitem{Fernandez-Moroni:2021nap}
G.~Fernandez-Moroni, R.~Harnik, P.~A.~N. Machado, I.~Martinez-Soler, Y.~F.
  Perez-Gonzalez, D.~Rodrigues et~al., \emph{{The physics potential of a
  reactor neutrino experiment with Skipper-CCDs: searching for new physics with
  light mediators}}, \href{https://doi.org/10.1007/JHEP02(2022)127}{\emph{JHEP}
  {\bfseries 02} (2022) 127},
  [\href{https://arxiv.org/abs/2108.07310}{{\ttfamily 2108.07310}}].

\bibitem{Drlica-Wagner:2020wck}
A.~Drlica-Wagner, E.~M. Villalpando, J.~O'Neil, J.~Estrada, S.~Holland,
  N.~Kurinsky et~al., \emph{{Characterization of skipper CCDs for cosmological
  applications}}, \href{https://doi.org/10.1117/12.2562403}{\emph{Proc. SPIE
  Int. Soc. Opt. Eng.} {\bfseries 11454} (2020) 114541A},
  [\href{https://arxiv.org/abs/2103.07527}{{\ttfamily 2103.07527}}].

\bibitem{DESI:2022lza}
{\scshape DESI} collaboration, D.~J. Schlegel et~al., \emph{{A Spectroscopic
  Road Map for Cosmic Frontier: DESI, DESI-II, Stage-5}},
  \href{https://arxiv.org/abs/2209.03585}{{\ttfamily 2209.03585}}.

\bibitem{arnaud2011handbook}
K.~Arnaud, R.~Smith and A.~Siemiginowska, \emph{Handbook of X-ray Astronomy},
  vol.~7.
\newblock Cambridge University Press, 2011.

\bibitem{uttley2014x}
P.~Uttley, E.~Cackett, A.~Fabian, E.~Kara and D.~Wilkins, \emph{X-ray
  reverberation around accreting black holes}, {\emph{The Astronomy and
  Astrophysics Review} {\bfseries 22} (2014) 1--66}.

\bibitem{Ingram:2019mna}
A.~Ingram and S.~Motta, \emph{{A review of quasi-periodic oscillations from
  black hole X-ray binaries: observation and theory}},
  \href{https://doi.org/10.1016/j.newar.2020.101524}{\emph{New Astron. Rev.}
  {\bfseries 85} (2019) 101524},
  [\href{https://arxiv.org/abs/2001.08758}{{\ttfamily 2001.08758}}].

\bibitem{CACKETT2021102557}
E.~M. Cackett, M.~C. Bentz and E.~Kara, \emph{Reverberation mapping of active
  galactic nuclei: From x-ray corona to dusty torus},
  \href{https://doi.org/https://doi.org/10.1016/j.isci.2021.102557}{\emph{iScience}
  {\bfseries 24} (2021) 102557}.

\bibitem{Mushotzky:2019lpm}
R.~F. Mushotzky et~al., \emph{{The Advanced X-ray Imaging Satellite}},
  {\emph{Bull. Am. Astron. Soc.} {\bfseries 51} (2019) 107},
  [\href{https://arxiv.org/abs/1903.04083}{{\ttfamily 1903.04083}}].

\bibitem{gaskin2019lynx}
J.~A. Gaskin, D.~A. Swartz, A.~Vikhlinin, F.~{\"O}zel, K.~E. Gelmis, J.~W.
  Arenberg et~al., \emph{Lynx x-ray observatory: an overview}, {\emph{Journal
  of Astronomical Telescopes, Instruments, and Systems} {\bfseries 5} (2019)
  021001--021001}.

\bibitem{Feigelson2022}
E.~D. Feigelson, V.~L. Kashyap and A.~Siemiginowska, \emph{Time Domain Methods
  for X-Ray and Gamma-Ray Astronomy}, pp.~1--26.
\newblock Springer Nature Singapore, Singapore, 2022.
\newblock 10.1007/978-981-16-4544-0135-1.

\bibitem{doi:10.1146/annurev-astro-052920-112338}
A.~Philippov and M.~Kramer, \emph{Pulsar magnetospheres and their radiation},
  \href{https://doi.org/10.1146/annurev-astro-052920-112338}{\emph{Annual
  Review of Astronomy and Astrophysics} {\bfseries 60} (2022) 495--558}.

\bibitem{boyle1974buried}
W.~Boyle and G.~Smith, \emph{Buried channel charge coupled devices},  Feb.~12,
  1974.

\bibitem{Du:2023soy}
P.~Du, D.~Ega\~na Ugrinovic, R.~Essig and M.~Sholapurkar, \emph{{Low-Energy
  Radiative Backgrounds in CCD-Based Dark-Matter Detectors}},
  \href{https://arxiv.org/abs/2310.03068}{{\ttfamily 2310.03068}}.

\bibitem{hynecek1979virtual}
J.~Hynecek, \emph{Virtual phase ccd technology},  in \emph{1979 International
  Electron Devices Meeting}, pp.~611--614, IEEE, 1979.

\bibitem{saks1980technique}
N.~Saks, \emph{A technique for suppressing dark current generated by interface
  states in buried channel ccd imagers}, {\emph{IEEE Electron Device Letters}
  {\bfseries 1} (1980) 131--133}.

\bibitem{ranuarez2006review}
J.~C. Ranu{\'a}rez, M.~J. Deen and C.-H. Chen, \emph{A review of gate tunneling
  current in mos devices}, {\emph{Microelectronics reliability} {\bfseries 46}
  (2006) 1939--1956}.

\bibitem{PhysRev.87.835}
W.~Shockley and W.~T. Read, \emph{Statistics of the recombinations of holes and
  electrons}, \href{https://doi.org/10.1103/PhysRev.87.835}{\emph{Phys. Rev.}
  {\bfseries 87} (Sep, 1952) 835--842}.

\bibitem{lenzlinger1969fowler}
M.~Lenzlinger and E.~Snow, \emph{Fowler-nordheim tunneling into thermally grown
  sio2}, {\emph{Journal of Applied physics} {\bfseries 40} (1969) 278--283}.

\bibitem{weinberg1982tunneling}
Z.~Weinberg, \emph{On tunneling in metal-oxide-silicon structures},
  {\emph{Journal of Applied Physics} {\bfseries 53} (1982) 5052--5056}.

\bibitem{maserjian1974tunneling}
J.~Maserjian, \emph{Tunneling in thin mos structures}, {\emph{Journal of Vacuum
  Science and Technology} {\bfseries 11} (1974) 996--1003}.

\bibitem{srivastava1985electrical}
J.~Srivastava, M.~Prasad and J.~Wagner, \emph{Electrical conductivity of
  silicon dioxide thermally grown on silicon}, {\emph{Journal of The
  Electrochemical Society} {\bfseries 132} (1985) 955}.

\bibitem{jerram2001llccd}
P.~Jerram, P.~J. Pool, R.~Bell, D.~J. Burt, S.~Bowring, S.~Spencer et~al.,
  \emph{The llccd: low-light imaging without the need for an intensifier},  in
  \emph{Sensors and Camera Systems for Scientific, Industrial, and Digital
  Photography Applications II}, vol.~4306, pp.~178--186, SPIE, 2001.

\bibitem{daigle2010darkest}
O.~Daigle, P.-O. Quirion and S.~Lessard, \emph{The darkest emccd ever},  in
  \emph{High Energy, Optical, and Infrared Detectors for Astronomy IV},
  vol.~7742, pp.~28--38, SPIE, 2010.

\bibitem{daigle2012characterization}
O.~Daigle, O.~Djazovski, D.~Laurin, R.~Doyon and {\'E}.~Artigau,
  \emph{Characterization results of emccds for extreme low-light imaging},  in
  \emph{High Energy, Optical, and Infrared Detectors for Astronomy V},
  vol.~8453, pp.~10--18, SPIE, 2012.

\bibitem{bush2021measurement}
N.~Bush, J.~Heymes, D.~Hall, A.~Holland and D.~Jordan, \emph{Measurement and
  optimization of clock-induced charge in electron multiplying charge-coupled
  devices}, {\emph{Journal of Astronomical Telescopes, Instruments, and
  Systems} {\bfseries 7} (2021) 016002--016002}.

\bibitem{sequin1974charge}
C.~Sequin, F.~Morris, T.~A. Shankoff, M.~Tompsett and E.~Zimany,
  \emph{Charge-coupled area image sensor using three levels of polysilicon},
  {\emph{IEEE Transactions on Electron Devices} {\bfseries 21} (1974)
  712--720}.

\bibitem{holland1989fabrication}
S.~Holland, \emph{Fabrication of detectors and transistors on high-resistivity
  silicon}, {\emph{Nucl. Instrum. Methods Phys. Res., Sect. A.;(Netherlands)}
  {\bfseries 275} (1989) }.

\bibitem{holland1996200}
S.~Holland, G.~Goldhaber, D.~E. Groom, W.~Moses, C.~Pennypacker, S.~Perlmutter
  et~al., \emph{A 200/spl times/200 ccd image sensor fabricated on
  high-resistivity silicon},  in \emph{International Electron Devices Meeting.
  Technical Digest}, pp.~911--914, IEEE, 1996.

\bibitem{holland2003fully}
S.~E. Holland, D.~E. Groom, N.~P. Palaio, R.~J. Stover and M.~Wei, \emph{Fully
  depleted, back-illuminated charge-coupled devices fabricated on
  high-resistivity silicon}, {\emph{IEEE Transactions on Electron Devices}
  {\bfseries 50} (2003) 225--238}.

\bibitem{spratt1997effects}
J.~Spratt, B.~Passenheim and R.~Leadon, \emph{The effects of nuclear radiation
  on p-channel ccd imagers},  in \emph{1997 IEEE Radiation Effects Data
  Workshop NSREC Snowmass 1997. Workshop Record Held in conjunction with IEEE
  Nuclear and Space Radiation Effects Conference}, pp.~116--121, IEEE, 1997.

\bibitem{hopkinson1999proton}
G.~Hopkinson, \emph{Proton damage effects on p-channel ccds}, {\emph{IEEE
  Transactions on Nuclear Science} {\bfseries 46} (1999) 1790--1796}.

\bibitem{bebek2002proton}
C.~Bebek, D.~Groom, S.~Holland, A.~Karcher, W.~Kolbe, J.~Lee et~al.,
  \emph{Proton radiation damage in p-channel ccds fabricated on
  high-resistivity silicon}, {\emph{IEEE Transactions on Nuclear Science}
  {\bfseries 49} (2002) 1221--1225}.

\bibitem{palik1998handbook}
E.~D. Palik, \emph{Handbook of optical constants of solids}, vol.~3.
\newblock Academic press, 1998.

\bibitem{hynecek1981virtual}
J.~Hynecek, \emph{Virtual phase technology: A new approach to fabrication of
  large-area ccd's}, {\emph{IEEE Transactions on Electron Devices} {\bfseries
  28} (1981) 483--489}.

\bibitem{69907}
B.~Burke and S.~Gajar, \emph{Dynamic suppression of interface-state dark
  current in buried-channel ccds},
  \href{https://doi.org/10.1109/16.69907}{\emph{IEEE Transactions on Electron
  Devices} {\bfseries 38} (1991) 285--290}.

\bibitem{de2022search}
C.~De~Dominicis, \emph{Search for light dark matter with DAMIC-M experiment},
  Ph.D. thesis, Ecole nationale sup{\'e}rieure Mines-T{\'e}l{\'e}com
  Atlantique, 2022.

\bibitem{GEANT4:2002zbu}
{\scshape GEANT4} collaboration, S.~Agostinelli et~al., \emph{{GEANT4--a
  simulation toolkit}},
  \href{https://doi.org/10.1016/S0168-9002(03)01368-8}{\emph{Nucl. Instrum.
  Meth. A} {\bfseries 506} (2003) 250--303}.

\bibitem{aguilar2022oscura}
A.~Aguilar-Arevalo, F.~A. Bessia, N.~Avalos, D.~Baxter, X.~Bertou, C.~Bonifazi
  et~al., \emph{The oscura experiment}, {\emph{arXiv preprint arXiv:2202.10518}
  (2022) }.

\bibitem{Essig:2011nj}
R.~Essig, J.~Mardon and T.~Volansky, \emph{{Direct Detection of Sub-GeV Dark
  Matter}}, \href{https://doi.org/10.1103/PhysRevD.85.076007}{\emph{Phys. Rev.
  D} {\bfseries 85} (2012) 076007},
  [\href{https://arxiv.org/abs/1108.5383}{{\ttfamily 1108.5383}}].

\bibitem{Dreyer:2023ovn}
C.~E. Dreyer, R.~Essig, M.~Fernandez-Serra, A.~Singal and C.~Zhen, \emph{{Fully
  ab-initio all-electron calculation of dark matter--electron scattering in
  crystals with evaluation of systematic uncertainties}},
  \href{https://arxiv.org/abs/2306.14944}{{\ttfamily 2306.14944}}.

\bibitem{SENSEI:2023zdf}
{\scshape SENSEI} collaboration, P.~Adari et~al., \emph{{SENSEI: First
  Direct-Detection Results on sub-GeV Dark Matter from SENSEI at SNOLAB}},
  \href{https://arxiv.org/abs/2312.13342}{{\ttfamily 2312.13342}}.

\bibitem{DAMIC-M:2023hgj}
{\scshape DAMIC-M} collaboration, I.~Arnquist et~al., \emph{{Search for Daily
  Modulation of MeV Dark Matter Signals with DAMIC-M}},
  \href{https://arxiv.org/abs/2307.07251}{{\ttfamily 2307.07251}}.

\bibitem{stover1997characterization}
R.~J. Stover, M.~Wei, Y.~J. Lee, D.~K. Gilmore, S.~E. Holland, D.~E. Groom
  et~al., \emph{Characterization of a fully depleted ccd on high-resistivity
  silicon},  in \emph{Solid State Sensor Arrays: Development and Applications},
  vol.~3019, pp.~183--188, SPIE, 1997.

\bibitem{janesick1990new}
J.~R. Janesick, T.~S. Elliott, A.~Dingiziam, R.~A. Bredthauer, C.~E. Chandler,
  J.~A. Westphal et~al., \emph{New advancements in charge-coupled device
  technology: subelectron noise and 4096 x 4096 pixel ccds},  in
  \emph{Charge-Coupled Devices and Solid State Optical Sensors}, vol.~1242,
  pp.~223--237, SPIE, 1990.

\bibitem{1050535}
D.~Wen, \emph{Design and operation of a floating gate amplifier},
  \href{https://doi.org/10.1109/JSSC.1974.1050535}{\emph{IEEE Journal of
  Solid-State Circuits} {\bfseries 9} (1974) 410--414}.

\bibitem{chandler1990sub}
C.~E. Chandler, R.~A. Bredthauer, J.~R. Janesick and J.~A. Westphal,
  \emph{Sub-electron noise charge-coupled devices},  in \emph{Charge-Coupled
  Devices and Solid State Optical Sensors}, vol.~1242, pp.~238--251, SPIE,
  1990.

\bibitem{10.1088/978-0-7503-3235-4}
K.~D. Stefanov, \emph{CMOS Image Sensors}.
\newblock 2053-2563. IOP Publishing, 2022,
  \href{https://doi.org/10.1088/978-0-7503-3235-4}{10.1088/978-0-7503-3235-4}.

\end{thebibliography}\endgroup

\end{document}